
\def\newpage{\vfill\eject}
\def\centreline{\centerline}

\def\spose#1{\hbox to 0pt{#1\hss}}
\def\simlt{\mathrel{\spose{\lower 3pt\hbox{$\mathchar"218$}}
     \raise 2.0pt\hbox{$\mathchar"13C$}}}
\def\simgt{\mathrel{\spose{\lower 3pt\hbox{$\mathchar"218$}}
     \raise 2.0pt\hbox{$\mathchar"13E$}}}
\def\today{\ifcase\month\or January\or February\or March\or April\or May\or
      June\or July\or August\or September\or October\or November\or December\fi
      \space\number\day, \number\year}

\def\ref{\par \noindent \hangindent=3pc \hangafter=1}

\def\om{$\omega = 0$\a}

\def\ss{self-similar\a}

\def\double_under#1{\underline{\underline{#1}}}

\def\dda{\underline{\underline{A}}}
\def\ddb{\underline{\underline{B}}}
\def\ddc{\underline{\underline{C}}}
\def\ddv{\underline{\underline{V}}}
\def\ddp{\underline{\underline{P}}}
\def\ddt{\underline{\underline{T}}}
\def\ddM{\underline{\underline{M}}}
\def\ddpe{\underline{\underline{p}}}
\def\ddte{\underline{\underline{t}}}
\def\ddae{\underline{\underline{a}}}
\def\ddce{\underline{\underline{c}}}
\def\ddbe{\underline{\underline{b}}}
\def\ddve{\underline{\underline{v}}}
\def\pmb#1{\setbox0=\hbox{#1}%
\kern-.025em\copy0\kern-\wd0
\kern.05em\copy0\kern-\wd0
\kern-.025em\raise.0433em\box0}
\def\coms{\mag=1200 \hsize=15truecm \hoffset=1.0truecm \vskip 1.2truecm
\voffset=1.6truecm \vsize=21truecm \tabskip=1em plus 2.5em minus 0.5em
\parindent 1.0truecm \parskip=10pt}
\def\a{\ }
\def\b{\ \ }
\def\z{\b \ }

\def\ref{\par \noindent \hangindent=3pc \hangafter=1}
\def\AaA{A\& A}

\def\ApJ{ApJ}

\def\MN{MNRAS}

\def\etal{{\it et al.\thinspace}}
\def\eg{{\it e.g.\ }}
\def\etc{{\it etc.\ }}
\def\ie{{\it i.e.\ }}
\mathchardef\twiddle="2218

\def\multleft#1{\hbox to size{\vbox {\halign {\lft{##}\cr #1}}\hfill}\par}
\def\multright#1{\hbox to size{\vbox {\halign {\rt{##}\cr #1}}\hfill}\par}

\def\<{\thinspace}

%
%
%
%

\coms
\centreline{\bf EQUIANGULAR SPIRAL MODES OF POWER LAW DISKS}

\centreline{D. Lynden-Bell and J.P.S. Lemos}
\centreline{Institute of Astronomy, The Observatories, Cambridge CB3 0HA}

\vskip 0.8truecm
The stability of \ss power-law disks to non-axially symmetrical 
disturbances of zero frequency (\om) is discussed.  It is established that 
marginally unstable modes {\it either} have \om {\it or} consist of 
continua with modes of {\it all} frequencies becoming unstable together.  A 
careful study of all modes close to \om yields a remarkable conclusion.

\newpage
\noindent
{\bf 1. Introduction.}

Following early developments by Brodetsky \& Snow, Kalnajs has given a 
pretty development of the theory of gravitational potential of equiangular 
spiral waves.  Donner, in a fine but unpublished Cambridge thesis (
supervised by DLB), gave an account of viscous spiral waves in both the 
linear and non-linear r\'egimes.  The relationship of the vanishing 
viscosity limit of these waves to the stability problem of spiral structure 
was not completely elucidated.  Zang and Toomre have explored the stability 
of the V = constant Mestel disk in stellar dynamics.  Zang found m = 1 more 
unstable than the m = 2 modes.  Toomre used the stable modes to illustrate 
very beautifully the effects of the swing amplifier.

Spruit in a more recent paper that parallels Donner's analysis has 
discussed the non-linear equiangular spiral viscous accretion disk.

We were led to take a new look at the inviscid problem because the 
dimensional argument of section 3 showed that there is no characteristic 
frequency of a `power-law' disk.  It was then unclear how such a disk could 
pick out a characteristic frequency for its marginally stable mode.  This 
paper finds all the \om modes and resolves this question.

\newpage
\noindent
{\bf 2. Self-Similarity.}

A flat disk is said to be \ss if any physical quantity $q$ defining the 
configuration at the point $R,\phi$ is related to that quantity at other 
points by
$$q(\lambda R, \phi + f(\lambda)) = A_1 (\lambda) q (R,\phi)\, .\eqno (2.1)$$
So the physical quantity at a point $\lambda$ times further out and rotated 
by $f(\lambda)$ is a multiple of that quantity at the original point.  When 
$\lambda = 1$ we ask that the transformed point be coincident with the 
original one so we take $f(1) = 0$ and $A_1(1) = 1$.  Furthermore, while 
relationship (1) shall be true of all the physical quantities defining the 
configuration and for all values of $\lambda$, nevertheless the function 
$A_1 (\lambda)$ will depend on the physical dimensions mass, length, time, 
\etc of $q$ and be such that $A_1(\lambda)\equiv 1$ if $q$ is dimensionless
.  $f(\lambda)$ does not depend on $q$.

Differentiating (2.1) with respect to $\lambda$ and then putting $\lambda = 
1$, we deduce
$$R {\partial q\over\partial R} + \alpha {\partial q\over\partial\phi} = 
b\, q\, ,\eqno (2.2)$$
where $\alpha = f'(1)$ and $b = A'_1(1)$.

Equation (2.2) may be rewritten
$$R {\partial Q\over\partial R} + \alpha\, {\partial Q\over\partial\phi} = 0\, ,
\eqno (2.3)$$
where 
$$Q = ln\, q - b\,ln\,R = ln (q\, R^{-b})\, .\eqno (2.4)$$

The general solution of (2.3) is
$$Q = S(\phi - \alpha\, ln\, R)\, ,\eqno (2.5)$$
where $S$ is an arbitrary function so the general solution of (2.2) is
$$q = R^b F(\phi - \alpha\, ln\, R)\, .\eqno (2.6)$$
Hence in all \ss disks the pressure, density, velocity of rotation, \etc 
are all of the form (2.6).  Notice that $S$ and $F$ are constant on 
equiangular spirals.
\vskip 0.6truecm
\noindent
{\bf 3. Axially-symmetrical \ss disks in Equilibrium}

When we impose axial symmetry the functions $F$ in equation (2.6) all 
become constants.  Our equilibrium disks will have a fluid velocity $V(R)
{\bf{\hat\phi}}$ and centrifugal force together with the pressure gradient in 
the disk will balance the gravity of the disk.  It is useful to define the 
fictitious circular velocity $V_c(R){\bf{\hat\phi}}$ as that velocity which 
would by itself balance the gravity $-g(R){\bf{\hat R}}$
$$+ {V_c^2\over R} = g(R)\, ,\eqno (3.1)$$
while
$$+ {V_0^2\over R} = g(R) + {1\over\Sigma_0} {\partial p_0\over\partial R}
\, ,\eqno (3.2)$$
where $\Sigma_0 (R)$ is the surface density of the disk and $p_0(R)$ is the 
integral of the pressure through the (negligible) thickness of the disk.  
In accordance with equation (2.6) we take $V_c$ of the form
$$V_c = B_1 R^{-\beta}\eqno (3.3)$$
and note that because $V_0$ and $V_c$ have the same dimensions, $V_0$ is 
also proportional to $R^{-\beta}$.  Now $p_0/\Sigma_0$ has the dimensions 
of $V_c^2$ so it must vary as $R^{-2\beta}$.  The dimensions of 
$G\,\Sigma$ are those of $[GM/R^2]$ and $g$, so $G\,\Sigma$ varies as $V_c^2/R 
\propto R^{-(2\beta+1)}$.

Putting these results together
$$V_0^2/R \propto V_c^2/R = g \propto G\,\Sigma \propto R^{-(2\beta+1)}\, , 
\eqno (3.4)$$
$$p_0 = \Sigma_0 \sigma^2 \propto R^{-(4\beta+1)}\, .\eqno (3.5)$$
Inserting these results into (3.2), we find
$$V_0^2 = V_c^2 [1 - (4\beta + 1) \sigma^2/V_c^2]\, .\eqno (3.6)$$
If the pressure decreases outwards $\sigma^2/V_c^2 \leq (4\beta+1)^{-1}$.

Now the gravitational fields of power-law disks are readily calculated \eg 
by Mestel's method, which gives
$$g = 2\pi G\, \Sigma_0/L(\beta)\, ,\eqno (3.7)$$
where
$$L(\beta) = {\Gamma(1-\beta) \Gamma({1\over2} + \beta)\over\Gamma (1+
\beta) \Gamma({1\over2} - \beta)} \z {\rm and}\a -{\textstyle{1\over2}} < 
\beta < {\textstyle{1\over2}}\, .\eqno (3.8)$$

We have found the approximate formula
$$L(\beta) = ({\textstyle{1\over2}} - \beta)(1.88\beta^3 + 1.78\beta^2 + 2.
44\beta + 2)/(2\beta + 1)\eqno (3.9)$$
gives a better than 1.2\% fit to this function in the required range 
$-{\textstyle{1\over2}} < \beta \leq {\textstyle{1\over2}}$ (Table 1).
\newpage
\centreline{\bf Table 1}
\bigskip
\settabs 3\columns
\+$\beta$&$L(\beta)$&Approximation\cr
\medskip
\hrule
\medskip
\+.5&0&0\cr
\+.4&.188&.188\cr
\+.3&.367&.368\cr
\+.2&.550&.552\cr
\+.1&.754&.755\cr
\+0&1.000&1.000\cr
\+-.1&1.326&1.329\cr
\+-.2&1.818&1.829\cr
\+-.3&2.726&2.755\cr
\+-.4&5.304&5.348\cr
\+-.5&$\infty$&$\infty$\cr
\medskip
\hrule

Hence
$$2\pi G\,\Sigma_0 = L(\beta) B^2_1 R^{-(2\beta+1)} = L(\beta)V_c^2/R\eqno (
3.10)$$
$$p_0 = \sigma^2 \Sigma_0 = \left({\sigma^2\over V_c^2}\right)\left({L\, 
B_1^4\over 2\pi G}\right) R^{-(4\beta+1)}\, .\eqno (3.11)$$
For given values of $B_1$ and $|\beta|<{\textstyle{1\over2}}$ and for a (
constant) chosen ratio of $\sigma^2/V_c^2\leq {1\over(4\beta+1)}$, equation 
(3.3), (3.6), (3.10) and (3.11) define an equilibrium \ss rotating disk.

\noindent
{$\underline{\hbox{The particular interest of the problem}}$}

Notice that the only dimensionful constants defining the equilibrium disks 
are $G = [M^{-1} L^3 T^{-2}]$ and $B_1 = [L^{\beta+1} T^{-1}]$.  The other 
constants involved $\beta$ and $\sigma/V_c$ are both dimensionless.  When 
we are interested in perturbations about the equilibrium we shall take 
perturbed surface pressures and perturbed surface densities to be related 
by
$$\Delta p/p_0 = \gamma\,\Delta\,\Sigma/\Sigma_0$$
where $\gamma$ is another dimensionless constant.  For $\beta\not= 1$ it is 
{\it not} possible to make a characteristic frequency from the properties 
of the equilibrium because nothing of dimensions $[T^{-1}]$ can be made 
from $G$ and $B_1$ which characterize the equilibrium.

Now consider an imaginary linear series of equilibria with the same values 
of $G$, $B_1$ and $\beta$ but with gradually decreasing values of 
$\sigma^2/V_c^2$ starting from the value $(4\beta+1)^{-1}$ which is entirely pressure 
supported.  We shall take $\gamma > 3/2$ so that the system is then stable. 
 As we decrease the pressure support and increase the rotation speed to 
compensate, we know we must encounter a system that is marginally unstable 
since the system with $\sigma = 0$ is unstable to Jeans's instability at 
very short wavelengths.  Furthermore, we could perform the experiment of 
exciting modes of only one chosen $e^{im\phi}$ symmetry.  For each $m$ 
value there will be a different marginally unstable mode.  Each of these 
must have a natural frequency $\omega_m$ which might or might not be zero.  
However, $\omega_m$ must be a property of the equilibrium configuration and 
that has no way of making a characteristic frequency since no constants 
with dimensions $T^{-1}$ can be made from $G$ and $B_1$.  We deduce that 
the characteristic frequencies of isolated marginally unstable modes, 
whatever the $m$ may be, must be zero.  This implies that the `pattern' 
speeds $\Omega_p = -\omega/m$ of isolated modes must also be zero.  As we 
shall see, this greatly simplifies the calculations.  However, there is 
another strange possibility originally envisaged by Birkhoff in his book.  
If a mode is not isolated but part of a continuum then it can be that the 
system first becomes unstable with a whole continuum of modes, each with a 
different frequency all becoming unstable together.  Such modes would be 
\ss copies of each other, that of frequency $\omega_m$ being related to that 
of frequency $\omega'_m = \lambda^{-(\beta+1)}\omega_m$ by having a scale 
in space $\lambda$ times as large.  In this case the system would first go 
unstable at all scales simultaneously as a result of its self similarity.  
Which of these possibilities occurs?

For axially-symmetrical modes instability sets in through $\omega = 0$ for 
all inviscid systems with no meridional motions (see \eg Lynden-Bell \& 
Ostriker).  Because of this first, Schmitz (1986, 1988 \& 1989) and Schmitz 
\& Ebert (1987), and later we (Lemos \etal\a 1991) studied just those modes 
and found that the stability criterion was remarkably close to Toomre's 
local one.  Furthermore, the radial displacements of the marginally stable 
modes were of the form
$$\xi_R \propto R^{(4\beta+1)/2} \cos (\alpha\, ln\, R + {\rm const})$$
or in complex form, $\xi_R \propto R^Z$ where $Z = 2\beta + {\textstyle
{1\over2}} + i\, \alpha$.  For non-axially symmetrical modes there is no 
such nutcracker theorem to indicate that instability should set in at \om 
but the above arguments indicate that either \om or the mode structure is 
truly remarkable with modes of all real frequencies going unstable 
simultaneously.  To spiral structure theorists it would be odd if marginal 
stability took place at \om for modes of every non-zero $m$, since this 
would imply a zero pattern speed for the spiral structure; however, the 
pattern speed is often associated with a corotation point at or near the 
edge of the disk and the power law disks are infinite so this could imply a 
zero pattern speed for {\it them} even if this were not generally the case.

These considerations greatly aroused our interest in studying modes with 
frequencies close \om for the power law disks.  As we shall show, the 
perturbation equations at \om are \ss under a radial scaling, so the \om 
modes can be found exactly for every $m$.  Having found all these modes we 
then embark on a more delicate discussion to discover whether such modes 
are marginally unstable or whether they are ordinary stable modes that 
happen to have zero frequency in non-rotating axes.  If the former were 
true, we would find the marginally unstable modes so we could deduce the 
criterion for stability and the form of the instability at each value of 
$m$.  If the latter were true, we would have proved the remarkable result 
that modes at all frequencies become marginally unstable together so these 
could be sought at any chosen frequency.  We shall adopt the Agatha 
Christie approach, not giving the answer until we find it.  Those spoil-
sports who merely want the culprit named can read the end of the paper 
first.

\vskip 0.6truecm
\noindent
{\bf 4. Disturbances with \om}

Many years ago now, Lynden-Bell \& Ostriker created a powerful formalism 
for discussing stability problems of this type.  They showed that the 
eigenvalue problem for $\omega$ would be reduced to the solution of the 
equation
$$-\omega^2 \dda\cdot{\bf\xi} + \omega \ddb\cdot{\bf\xi} + \ddc\cdot{\bf\xi} 
= 0\eqno (4.1)$$
where $\dda$, $\ddb$ and $\ddc$ were Hermitian operators and ${\bf\xi}e^{i
\omega t}$ was the displacement vector involved in the perturbation.  
$\dda$ is positive definite.  For the particular case in which the 
undisturbed equilibrium is axially symmetrical with motion in the ${\bf{\hat
\phi}}$ direction only (\ie no meridional circulation), ${\bf\xi}$ may be 
taken to vary as $e^{im\phi}$ and the different $m$ components each obey a 
similar equation with simpler operators $\dda_m$, $\ddb_m$ and $\ddc_m$ 
replacing $\dda$, $\ddb$, $\ddc$.  Lynden-Bell and Ostriker's formalism was
developed for a 3-dimensional system of density $\rho_0({\bf r})$ so we 
must integrate in the vertical direction over all $z$ to obtain equations 
for a flat disk in which case our $\Sigma_0(R)$ replaces $\rho_0({\bf r})$
and all $z$ components vanish.  The operator equation then takes the simple 
form
$$\left(-\omega^2 \dda_m + \omega\ddb_m + \ddc_m\right)\cdot\pmatrix{\xi_R&
+&i\xi_\phi\cr
\xi_R&-&i\xi_\phi\cr} = 0\eqno (4.2)$$
where
$$\dda_m = \Sigma_0 \pmatrix{1&0\cr
0&1\cr} \eqno (4.3)$$
$$\ddb_m = - 2\Sigma_0 \Omega_0 \pmatrix{m+1&0\cr
0&m-1\cr} \eqno (4.4)$$
where $\Omega_0(R) = V_0(R)/R$,
$$\ddc_m = \ddt_m + \ddv_m + \ddp_m\, .\eqno (4.5)$$
Here
$$\ddt_m = -\Sigma_0 \Omega_0^2 \pmatrix{(m+1)^2&0\cr
0&(m-1)^2\cr} \eqno (4.6)$$
while $\ddv_m$ and $\ddp_m$ are the operators obtained from $\ddv \cdot{\bf
\xi}$ and $\ddp \cdot{\bf\xi}$ by taking ${\bf\xi}$ to behave as $e^{im
\phi}$ and making them operate on the vector ${\bf\zeta} = \pmatrix{\xi_R&+
&i\xi_\phi\cr
\xi_R&-&i\xi_\phi\cr}$ instead of $\pmatrix{\xi_R\cr\xi_\phi\cr}$.

$\ddv\cdot{\bf\xi}$ and $\ddp\cdot{\bf\xi}$ are defined by
$$-\ddv\cdot{\bf\xi} = \Sigma_0 \Delta({\bf\nabla}\psi)\eqno (4.7)$$
$$\ddp\cdot{\bf\xi} = \Sigma_0 \Delta \left({1\over\Sigma} {\bf\nabla} p\right)
\eqno (4.8)$$
where $\psi$ is the gravitational potential and $\Delta$ is the Lagrangian 
displacement operator
$$\Delta \left[Q({\bf r},t)\right] = Q({\bf r} + {\bf\xi}, t) - Q_0({\bf r}
,t)\, .\eqno (4.9)$$
Here $Q$ is any quantity evaluated in the perturbed flow, $Q_0$ is the same 
quantity evaluated in the unperturbed flow and ${\bf\xi} = {\bf\xi}({\bf r}
,t)$.  To first order in ${\bf\xi}$, $\Delta$ is related to the Eulerian 
change $\delta$ evaluated at the same point in both flows by
$$\Delta = \delta + {\bf\xi}\cdot{\bf\nabla}\, .\eqno (4.10)$$
Lynden-Bell \& Ostriker (equation 25) show
$$\ddp\cdot{\bf\xi} = {\bf\nabla}\left[(1-\gamma)p_0{\rm div}{\bf\xi}\right] -
p_0 {\bf\nabla}({\rm div}{\bf\xi}) - {\bf\nabla}\left[({\bf\xi}\cdot{\bf\nabla}
)p_0\right] + ({\bf\xi}\cdot{\bf\nabla}){\bf\nabla}p_0\eqno (4.11)$$
and from (4.7) and (4.10)
$$-\ddv\cdot{\bf\xi} = \Sigma_0({\bf\xi}\cdot{\bf\nabla}){\bf\nabla}\psi_0 
+ \Sigma_0 {\bf\nabla} \delta\psi\, .\eqno (4.12)$$
$\delta\psi$ has to be evaluated in terms of ${\bf\xi}$ by use of Poisson's 
integral remembering that $\delta\Sigma = -{\rm div}(\Sigma_0{\bf\xi})$.  
However, although Lynden-Bell \& Ostriker used that method to demonstrate 
that $\ddv$ was Hermitian, there is a simpler method for the \om modes 
based on their self-similarity.

The scalars involved in the variational principle of Lynden-Bell \& 
Ostriker are quantities such as $\int\int{\bf\zeta^*}\cdot\ddt_m\cdot\zeta
Rd\phi dR$. The operator $T_m$ itself is proportional to $\Sigma_0 
\Omega_0^2 \propto R^{-(4\beta+3)}$ while the $RdR$ integrations raise this 
to $R^{-(4\gamma+1)}$.  This suggests that $\zeta$ will vary as $R^{+(4
\beta+1)/2}$, so we shall hereafter write
$${\bf\zeta} = R^{(4\beta+1)/2}{\bf\eta}\, .\eqno (4.13)$$
This agrees with the behaviour found for the axially symmetrical modes by 
Lemos \etal\a who discuss the interesting behaviour at large $R$.  We write 
$D = R \partial/\partial R$ and after some lengthy calculations done independently 
by each of us, we obtain
$$\ddp_m\cdot {\bf\zeta} = R^{2\beta-3/2} \Sigma_0 V_c^2 \ddpe_m\cdot {\bf\eta} 
\eqno (4.14)$$
where
$\ddpe_m\cdot{\bf\eta} = -{\textstyle{1\over2}}\sigma^2_0 \ddM \cdot 
\pmatrix{\eta_+\cr
\eta_-\cr}$ and $\ddM$ is the matrix
$$\pmatrix{\gamma[D^2-(\beta_2+m)^2]+2\beta_1(1+m)&\gamma[(D-m)^2-\beta^2_2]\cr
\gamma[(D+m)^2-\beta^2_2]&\gamma[D^2-(\beta_2-m)^2]+2\beta_1(1-m)\cr}\eqno 
(4.15)$$
Here
$$\sigma^2_0 = \sigma^2/V_c^2\, , \z \beta_1 = 4\beta+1 \a{\rm and}\a
\beta_2 = 2\beta + 3/2\, .\eqno (4.16)$$
Similarly
$$\ddt_m\cdot{\bf\zeta} = R^{2\beta-3/2} \Sigma_0 V_0^2 \ddte_m\cdot {\bf 
\eta}\eqno (4.17)$$
where using (3.6) we find
$$\ddte_m\cdot{\bf\eta} = -(1-\beta_1 \sigma_0^2)\pmatrix{(m+1)^2&0\cr
0&(m-1)^2\cr}\cdot\pmatrix{\eta_+\cr
\eta_-\cr}\, .\eqno (4.18)$$

To calculate the gravity term we see from (4.12) that we need an expression 
for $\delta\psi$ which arises from $\delta\Sigma$.
$$\eqalign{\delta\Sigma&= -{\rm div}(\Sigma_0{\bf\xi}) = \Sigma_0 R^{-1}
(2\beta \xi_R - D\xi_R - im\xi_R\phi)\cr
&= -\Sigma_0 R^{2\beta-1/2}\left[\left(D+{\textstyle{1\over2}}\right)\eta_R
+ im\eta_\phi\right]\cr}\eqno (4.19)$$

Now $\Sigma_0 \propto R^{-2\beta-1}$, so
$$R^{3/2} \delta\, \Sigma \propto \left(D + {\textstyle{1\over2}}\right)
\eta_R + im\eta_\phi\, .\eqno (4.20)$$
Kalnajs has shown that for flat disks the Fourier transform with respect to 
$u = ln R$ of $R^{3/2} \delta\Sigma$ is simply related to the Fourier 
transform in $lnR$ of $R^{1/2} \delta\psi$.  In particular if $R^{3/2} 
\delta\Sigma \propto e^{iku} e^{im\phi}$, then
$$R^{1/2} \delta\psi = 2\pi\, G\, K(k,m) R^{3/2} \delta\Sigma \eqno (4.21)$$
where
$$K(k,m) = {\textstyle{1\over2}} {\Gamma (m+{\textstyle{1\over2}} + ik)/2)
\Gamma(m+{\textstyle{1\over2}} - ik)/2)\over \Gamma((m+{\textstyle{3
\over2}} + ik)/2) \Gamma((m+{\textstyle{3\over2}} - ik)/2)}\, .\eqno (4.22)$$
For $m\geq 2$ Donner has shown this to be well approximated by
$$K(k,m) \simeq 1/s\eqno (4.23)$$
where
$$s^2 = (k^2 + m^2 + {\textstyle{1\over4}})\, .\eqno (4.24)$$
To extend Donner's approximation to $m = 0,1$ we use the exact recurrence 
relation
$$K(k,m-1) K(k,m) = {1\over (m-{\textstyle{1\over2}})^2 + k^2}\eqno
(4.25)$$
which follows from the properties of the $\Gamma$ function.  We thus obtain
$$K(k,1) = \left[K(k,2)\left({\textstyle{9\over4}} + k^2\right)\right]^{-1}
\simeq \sqrt{k^2+{\textstyle{17\over4}}}\left(k^2+{\textstyle{9\over4}}
\right)^{-1}\eqno (4.26)$$
$$K(k,0) = \left[K(k,1)\left({\textstyle{1\over4}} + k^2\right)\right]^{-1}
\simeq \left[\sqrt{k^2 + {\textstyle{17\over4}}}\right]^{-1}
\left({k^2 + {\textstyle{9\over4}}\over k^2 + {\textstyle{1\over4}}}\right)
\, .\eqno (4.27)$$
These formulae are accurate to better than 1\% as Table 2 shows.  Thus when
${\bf\eta} \propto e^{iku+im\phi}$, we find using (4.20) and (4.21)
$$R^{1\over2} \delta\psi = 2\pi\, G\, K \, R^{3\over2}\delta\Sigma =
-2\pi\, G\, K\, \Sigma_0 R^{2\beta+1}[(ik + {\textstyle{1\over2}}\eta_R
+ im\eta_\phi]\, .\eqno (4.28)$$
(More generally a superposition of components of different $k s$ may be 
needed).

Returning to (4.12) and using (3.10), we finally obtain
$$\ddv_m\cdot\xi - \Sigma_0 V_c^2 R^{2\beta=3/2} \ddve_m\cdot {\bf\eta}
\eqno (4.29)$$
where (when only one $k$ component is present)
$$\ddve_m\cdot {\bf\eta} = \left[-{\textstyle{1\over2}}L\, K
\pmatrix{k^2 + \left({\textstyle{1\over2}}+m\right)^2&{\textstyle{1\over4}}
-(ik-m)^2\cr
{\textstyle{1\over4}}-(ik+m)^2&k^2+\left({\textstyle{1\over2}}-m\right)^2
\cr}-\pmatrix{\beta&\beta+1\cr
\beta+1&\beta\cr}\right]\cdot\pmatrix{\eta_+\cr
\eta_-\cr}\eqno (4.30)$$
Our operator equation (4.2) can now be written in dimensionless form
$$\left(-{\omega^2\over\Omega_c^2} \ddae_m + {\omega\over\Omega_c} 
\ddbe_m + \ddce_m\right) \cdot {\bf\eta} = 0\, ,\eqno (4.31)$$
where $\Omega_c = V_c/R$,
$$a_m = \pmatrix{1&0\cr 0&1\cr}\, ,\eqno (4.32)$$
$$b_m = -{2\Omega_0\over\Omega_c} \pmatrix{m+1&0\cr 0&m-1\cr}\, ,\eqno
(4.33)$$ and
$$\ddce_m = \ddte_m + \ddpe_m + \ddve_m\, .\eqno (4.34)$$
Notice that when \om only the $\ddce_m$ term survives.  No term in 
$\ddte_m$, $\ddpe_m$ or $\ddve_m$ involves $u=lnR$ except through the 
operator $D=\partial/\partial u$.  Hence these modes have an $e^{iku}$ 
behaviour and we may replace the operator $D$ by $ik$ where it occurs in
$\ddpe_m$.  This justifies our use of this form for $\ddve_m$ for these 
modes and reduces the problem of finding them to algebra.  To find such \om 
modes we need the determinant of $\ddce_m$ to be zero.  We find them by 
calculating the determinant numerically as a function of $k$.  Figure 1 
plots the $k$ so found against $\beta$.  Figure 2 shows the spirals that 
result.  Toomre's $Q$ parameter 
gives the stability of the axially-symmetric modes, both in the local 
approximation (and also quite accurately, $<4$\%, for the global modes of 
Lemos \etal).  To compare with the Local analysis of a gas disk we recall 
that for such modes
$$\omega^2 = k^2_1 c^2 - 2\pi\, G\, \Sigma_0 |k_1| + \kappa^2\eqno (4.35)$$
where we use the suffix 1 to distinguish the wave number $k_1$ in $R$ from 
the wave number $k$ in $lnR$ used earlier.  $c$ is the velocity of sound 
$(\gamma p_0/\Sigma_0)^{1\over2}$, $\kappa$ is related to $\Omega_c$ via
$$\kappa^2 = 4\Omega_c \left[{1\over 2R} {d\over dR}(\Omega_c R^2)\right]
= 2(1-\beta) \Omega_c^2\, .\eqno (4.36)$$
The expression for $\omega^2$ may be rewritten
$$\omega^2 = c^2 \left(k_1 - {\pi\, G\, \Sigma_0\over c^2}\right)^2 +
(\kappa^2 - \pi^2 G^2 \Sigma_0^2/c^2)\eqno (4.37)$$
which demonstrates that the least stable wave number is given by $k_1 =
\pi\, G\, \Sigma_0/c^2$ and the condition for stability to axially 
symmetrical modes is $Q > 1$, where
$$Q = {\kappa c\over\pi\, G\, \Sigma_0} = {\sqrt{8(1-\beta)}\over 
L(\beta)} {c\over V_c} = {\sqrt{8\gamma(1-\beta)}\over L(\beta)}\, 
\sigma_0\, .\eqno (4.38)$$

It is well known that Toomre's original criterion for stars only differs 
from this in that 3.36 replaces the value of $\pi$ in the denominator.  
Here the gaseous $Q$ defined above is the relevant one.  It is also 
interesting to compare the least stable wavenumber $k'_0 = k_0R$ with the value of 
$k$.  From (4.35)
$$\eqalign{k_0R &= \pi\, G\, \Sigma_0\, R/c^2 = {\textstyle{1\over2}}
L(\beta) V_c^2/c^2 = {\textstyle{1\over2}} L(\beta)/(\gamma\sigma_0^2)\cr
k_0R &= {4(1-\beta)\over Q^2 L(\beta)}\cr}\, .\eqno (4.39)$$

When the modes are non-axially symmetrical, the approximation commonly used 
for tightly wound modes is found from (4.35) or (4.37) by replacing 
$\omega$ by $\omega + m \Omega_0(R)$.  Equation (4.35) then takes the form
$$-{\omega^2\over\Omega_c^2} - {\omega\over\Omega_c} 2m {\Omega_0\over
\Omega_c} + \left\{m^2(1-\beta_1\sigma_0^2) - [k_1^2 c^2-2\pi \, G\, 
\Sigma_0 |k| + \kappa^2]{1\over \Omega_c^2}\right\} = 0$$
apart from the matrix form of equation (4.31) we see there is a close 
similarity with this approximate dispersion relationship.  We therefore 
expect that \om will occur near where the large curly bracket is zero and 
that the value of $k$ for which \om will be approximately
$$k'_1 = k_1R = \left[m^2 (1-(4\beta+1)\sigma_0^2 - {\kappa^2\over
\Omega_c^2} + {\pi^2 G^2 \Sigma_0^2\over\Omega_c^2 c^2}\right]^{1\over2}
{V_c\over c} + {\pi\, G\, \Sigma_0 R\over c^2}$$
Re-writing this expression in terms of our dimensionless variables we find
$$k'_1 = \left[m^2(1-(4\beta+1)\sigma_0^2) - 2(1-\beta) + {L(\beta)^2
\over 4\gamma\sigma_0^2}\right]^{1\over2} {1\over\gamma^{1\over2}\sigma_0}
+ {L(\beta)\over 2\gamma\sigma_0^2}\eqno (4.40)$$
which is compared with the exactly calculated results in Figure 3.  $k'_1$
given by (4.40) is good approximation for the \om modes in all but the very 
open spirals.  $k_0R$ shows no agreement.

\vskip 0.6truecm
\noindent
{\bf 5. Marginal Instability}

The existence of stationary modes does not imply marginal stability.  To 
ensure the latter we need the limit of growing modes as the growth rate 
tends to zero.  Returning to (4.1) we see that unless $\ddb\cdot{\bf\xi}$ 
is zero, modes with $\omega$ close to zero will obey
$$\omega\, \ddb\cdot{\bf\xi} = - \ddc\cdot{\bf\xi}$$
However $\ddb$ and $\ddc$ are Hermitian and we can see from (4.33) that at 
least for all modes with $m > 1$, $\ddb$ is negative definite.  Hence the 
eigenvalues $\omega$ are necessarily real and for any eigen $\xi$
$$\omega = {-{<{\bf\xi}^*\cdot\ddc\cdot{\bf\xi}>\over<\xi^*\cdot\ddb\cdot
{\bf\xi}>}}$$
Thus, provided $<{\bf\xi}^*\cdot\ddb\cdot{\bf\xi}> \not= 0$, all the eigen 
frequencies close to \om are real.  Now we may evaluate
$<{\bf\xi}^*\cdot\ddb\cdot{\bf\xi}>$ for our \om mode it can only be 
slightly different for the infinitesimally different mode with $\omega
\rightarrow 0$.  Hence, provided $<{\bf\xi}^*\cdot\ddb\cdot{\bf\xi}> \not= 0$
for our \om modes, there can be no marginally unstable modes close to \om.

Now in the form (4.33) $b_m$ is diagonal, so all we need to ensure is that 
for our \om mode
$$\eta^*_+ b_{11}\eta_+ + \eta_-^* b_{22}\eta_- \not= 0$$
multiplying by $c_{22}$ the LHS becomes
$$\eta_+^* \eta_{_+} (b_{11} c_{22} + b_{22} c_{11})$$
where we have used the fact that $\ddce_m\cdot{\bf\eta} = 0$ to change the
$\eta_-$ terms to $\eta_+$ ones by writing
$$c_{22}\eta_-^* = -c_{12}\eta_+^* \a {\rm and}\a c_{12}\eta_- = -c_{11}
\eta_+\, .$$
Thus, proveded $\eta_+$ is non zero, the condition $b_{11}c_{22}+b_{22}
c_{11} \not= 0$ implies that $<{\bf\xi}\cdot\ddb\cdot{\bf\xi}> \not= 0$ for the 
marginal mode. (Even if $\eta_+$ is zero, a similar argument multiplying by 
$c_{11}$ to start with yields the same result provided $\eta_-$ is non zero 
and they can not both be zero as then $\eta$ vanishes).  Now
$$b_{11}c_{22}+b_{22}c_{11} = {2\Omega_0 m\over\Omega_c}\left[{-2(m^2-1+\beta)
-L\, K(m^2+k^2-{\textstyle{3\over4}})+\atop
+\sigma_0^2\left\{2\beta_1(m^2-1)+\gamma[(k-\beta_2)^2+m^2]\right\}}\right]$$
To find out whether this can ever be zero for a marginally stable mode we 
solved the equation $b_{11}c_{22}+b_{22}c_{11} = 0$ for $\sigma_0^2$ 
and inserted the result into $||c_m||$ for each $k$, the result was 
negative for every $k$ and every $m \not= 0$.

We deduce that any marginally stable mode with $m \not= 0$ has a non-zero 
frequency $\omega_m$ and an associated corotation resonance (provided 
$\omega_m$ is positive) at $R^{\beta
+1} = B_1/\omega_m$.  The re-scalings of this mode by factors $\lambda$ 
yield an infinity of marginally stable modes with frequencies 
$\lambda^{-(\beta+1)}\omega_m$.  Thus, marginally stable modes occur 
simultaneously at all frequencies.  We find this result to be unexpected 
and remarkable.  One may now search for a marginally unstable mode by first 
choosing a frequency; any one is as good as any other; it is 
unnecessary to search through all possible frequency as one would 
have to do for a non-self-similar system.

\noindent
{$\underline{\hbox{Axially Symmetric Modes}}$}

It is of interest to see how the \om modes studied earlier fit into the 
present analysis.  Setting $m = 0$, $\ddce_m$ becomes a multiple of the 
matrix $\pmatrix{1&1\cr 1&1\cr}$ and $b_{11}c_{22}+b_{22}c_{11}$ vanishes.  
Thus the determinant $||\ddce_m||$ and $<{\bf\xi}^*\cdot\ddb\cdot{\bf\xi}>$ both 
vanish.  This is associated with the existence of the trivial neutral 
displacement with $\xi_\phi$ independent of $\phi$.  To find the marginally 
unstable modes one must expand the determinant
$$\Bigl\arrowvert\Bigl\arrowvert - \left({\omega\over\Omega_c}\right)^2 \ddae_m +
{\omega\over\Omega_c} \ddbe_m + \ddce_m \Bigl\arrowvert\Bigl\arrowvert$$
to all orders in $\omega/\Omega_c$ when one recovers
$${\omega^4\over\Omega_c^4}||\ddae_m|| - \left({\omega\over\Omega_c}\right)
^3 (a_{11}b_{22}+a_{22}b_{11})-\left({\omega\over\Omega_c}\right)^2
\left({{a_{11}c_{22}+a_{22}c_{11}}\atop {-b_{11}b_{22}}}\right) = 0$$
but for $m = 0$, $a_{11}b_{22} + a_{22}b_{11} = 0$, so the equation for the 
non-trivial \om axially symmetric modes is
$${\omega^2\over\Omega_c^2} ||a_m|| = (a_{11}c_{22}+a_{22}c_{11}-b_{11}
b_{22}) = 0$$
For $m=0$, $a_{11}c_{22}+a_{22}c_{11}-b_{11}b_{22} = 0$ yields
$$\sigma_0^2[\gamma(k^2+\beta^2_2) - 4\beta_1] + 2(1-\beta) - LK(k^2 +
{\textstyle{1\over4})} = 0$$
which is the condition of marginal stability for $m = 0$ modes found by 
Lemos \etal\a in a slightly different notation.

\noindent
{\bf Conclusion}.

In line with the anti-spiral theorem of Lynden-Bell and Ostriker the 
condition $||c_m|| = 0$ involves $k^2$, $m^2$ and $\Omega^2_c$ but not the 
signs of $k$, $m$, $\Omega_c$.  This implies that to every trailing spiral 
mode with \om there corresponds a leading spiral mode with \om and that we 
could take pairs of linear combinations of them which are not spiral at all!  
Only for growing or decaying modes can this symmetry be broken in a 
dissipationless system.

The non-axially symmetric \om modes are not the key to the stability 
problem, rather they are examples of stable steady modes which can be found 
exactly.  However, that result leads to the remarkable conclusion that 
modes of {\it all frequencies} become marginally unstable together, so that 
marginally stable modes can be sought at any frequency and others can then 
be found by self-similar scaling.

\noindent
{\bf Acknowledgement}

We thank Dr J.F. Harper for suggesting the possibility of all modes of a 
self-similar class going unstable together and referring us to Birkhoff's 
book.

\centreline{\bf References}

\ref
Birkhoff, G. 1960 {\it Hydrodynamics: A study in logic, fact \& similitude}. 
Princeton U.P.
\ref
Brodetsky, S. 1917, {\it Q. Jl. Pure \& Appl. Math} {\bf 48}, 58.
\ref
Donner, K.J. 1979, Ph.D. Thesis, Cambridge University.
\ref
Kalnajs, A.J. 1971, \ApJ {\bf 166}, 275.
\ref
Kalnajs, A.J. 1976, \ApJ {\bf 205}, 751.
\ref
Lemos, J.P.S., Kalnajs, A.J. \& Lynden-Bell, D. 1991, \ApJ (June).
\ref
Lynden-Bell, D. \& Ostriker, J.P. 1967, \MN {\bf 136}, 243.
\ref
Lynden-Bell, D. \& Kalnajs, A.J. 1972, \MN {\bf 157}, 730.
\ref
Mestel, L. 1963, \MN {\bf 126}, 553.
\ref
Schmitz, F. 1986, \AaA {\bf 169}, 171.
\ref
Schmitz, F. 1988, \AaA {\bf 200}, 120.
\ref
Schmitz, F. 1990, \AaA {\bf 230}, 509.
\ref
Schmitz, F. \& Ebert, R. 1987, \AaA {\bf 181}, 41.
\ref
Snow, C. 1952, {\it Hypergeometric \& Legendre Functions with Applications 
to Integral Equations of Potential Theory}, NBS AMS-19, Washington.
\ref
Spruit, H.C. 1987, \AaA {\bf 184}, 173.
\ref
Toomre, A. 1964, \ApJ {\bf 139}, 1217.
\ref
Toomre, A. 1981, in {\it The Structure \& Evolution of Normal Galaxies}, 
eds. Fall, S.M. \& Lynden-Bell, D., Cambridge University Press, pp.111--
136.
\ref
Zang, T.A. 1976, Ph.D. Thesis, M.I.T. Cambridge, Mass.
\newpage
\centreline{\bf Table 2}
\bigskip
\vbox{\tabskip=0pt \offinterlineskip
\def\tablerule{\noalign{\hrule}}
\halign to \hsize{\strut#& \vrule#\tabskip=1.5em plus2em minus1em&
\hfil#\hfil&\vrule#&
\hfil#\hfil&\hfil#\hfil&\vrule#&
\hfil#\hfil&\hfil#\hfil&\vrule#&
\hfil#\hfil&\hfil#\hfil&\vrule#\tabskip=0pt\cr\tablerule
&&&&&&&&&&&&\cr
&&k&&K(k,0)&Approx&&K(k,1)&Approx&&K(k,2)&Approx&\cr
&&&&&&&&&&&&\cr
\tablerule
&&&&&&&&&&&&\cr
&&0 &&4.377&4.366&&0.914&0.916&&0.486&0.485&\cr
&&&&&&&&&&&&\cr
&&0.2&&3.834&3.812&&0.9018&0.904&&0.484&0.483&\cr
&&&&&&&&&&&&\cr
&&0.4&&2.810&2.799&&0.868&0.871&&0.478&0.476&\cr
&&&&&&&&&&&&\cr
&&0.6&&2.004&1.993&&0.818&0.823&&0.468&0.466&\cr
&&&&&&&&&&&&\cr
&&0.8&&1.479&1.468&&0.760&0.765&&0.456&0.452&\cr
&&&&&&&&&&&&\cr
&&1&&1.145&1.135&&0.699&0.705&&0.440&0.436&\cr
&&&&&&&&&&&&\cr
&&2&&0.519&0.512&&0.453&0.460&&0.353&0.348&\cr
&&&&&&&&&&&&\cr
&&4&&0.252&0.250&&0.244&0.247&&0.224&0.222&\cr
&&&&&&&&&&&&\cr
&&6&&0.167&0.166&&0.165&0.166&&0.159&0.158&\cr
&&&&&&&&&&&&\cr
&&8&&0.125&0.125&&0.124&0.125&&0.121&0.121&\cr
&&&&&&&&&&&&\cr
&&10&&0.1001&0.0999&&0.0996&0.0999&&0.0982&0.0979&\cr
&&&&&&&&&&&&\cr
\tablerule
}}
\newpage
\centreline{\bf Figure Captions}

\noindent
{\bf Figure 1}.  $k$ the dimensionless wavenumber in $lnR$ space for 
$\omega =0$ waves in power law disks with rotations $V \propto R^{-\beta}$. 
 The `ring' modes $m=0$ are marginally unstable.  The $m \not= 0$, $\omega 
= 0$ modes are stable.

\noindent
{\bf Figure 2a}. $\omega = 0$ modes of $V \propto R^{-\beta}$ gas disks: On 
the left $\beta = 0$, $Q = 1$, $\gamma = 2$; bottom $m = 1$, $k = 7$; 
second $m = 2$, $k = 9.45$; third $m = 3$, $k = 11.9$; top $m = 4$, $k = 
14.5$.  On the right $\beta = 0$, $Q = 1.5$, $\gamma = 2$; bottom $m = 1$, 
$k = 2.6$; second $m = 2$, $k = 4.6$; third $m = 3$, $k = 6.1$; top $m = 4$, 
$k = 7.7$.  Notice that the larger $Q$ gives more open spirals.

\noindent
{\bf Figure 2b}. $\beta = -0.2$.  For equal $Q$ and $m$ the less sheared 
rotation law gives more open spirals $Q = 1$, $\gamma = 2$, $m = 2$, $k = 
5.8$.

\noindent
{\bf Figure 3}. The dimensionless wavenumber of the exact $\omega = 0$ mode 
is plotted against $k'_0$, the value of $k_0 R$ for {\it marginal 
stability} in the local tightly-wound approximation.  $k$ and $k'_0$ give 
disparate curves $k'_1$, the $k_1 R$ of the $\omega = 0$ {\it stable} mode 
in the tightly wound approximation is a reasonably good approximation for 
large $k$.

\end